\documentclass[reprint,twocolumn,floatfix,preprintnumbers,prb ,nofootinbib,hyperref]{revtex4-2}
\usepackage[T2A,T1]{fontenc}
\usepackage{amsmath}
\usepackage{amssymb}
\usepackage[utf8]{inputenc}
\usepackage{graphicx}
\usepackage{makecell}
\usepackage{diagbox}
\usepackage{amssymb,amsmath}
\usepackage[dvipsnames]{xcolor}

\usepackage[breaklinks=true,hidelinks]{hyperref}
\hypersetup{allcolors=[rgb]{0.0 0.0 0.6},linkcolor=[rgb]{0.75 0.05 0.05}}

\newcommand{\om}{\omega}      
\newcommand{\kap}{\varkappa}      

\graphicspath{{./figures}}

\begin{document}
\title{\textbf{\normalsize{Tunable Wire Metamaterials for an Axion Haloscope}}}

\author{\normalsize{Nolan Kowitt$^\ast$$^\dagger$}}
\author{Dajie Sun$^\dagger$}
\author{Mackenzie Wooten}
\author{Alexander Droster}
\author{Karl van Bibber}
\affiliation{\emph{Department of Nuclear Engineering, University of California Berkeley}, 94720 \emph{USA}}
\author{\normalsize{Rustam Balafendiev$^\dagger$}}
\author{Maxim A. Gorlach}
\author{Pavel A. Belov}
\affiliation{\emph{ITMO University}, 197101 \emph{St. Petersburg, Russia}}
\date{\today}

\begin{abstract}
Metamaterials based on regular two-dimensional arrays of thin wires have attracted renewed attention in light of a recently proposed strategy to search for dark matter axions. When placed in the external magnetic field, such metamaterials facilitate resonant conversion of axions into plasmons near their plasma frequency. Since the axion mass is not known {\it a priori}, a practical way to tune the plasma frequency of metamaterial is required. In this work, we have studied a system of two interpenetrating rectangular wire lattices where their relative position is varied.  The plasma frequency as a function of their relative position in two dimensions has been mapped out experimentally, and compared with both a semi-analytic theory of wire-array metamaterials and numerical simulations.  Theory and simulation yield essentially identical results, which in turn are in excellent agreement with experimental data.  Over the range of translations studied, the plasma frequency can be tuned over a range of 16\%.
\end{abstract}

\maketitle

\footnotetext[1]{nkowitt@berkeley.edu}
\footnotetext[2]{These authors have equally contributed to the paper.}

\section{Introduction}

Metamaterials are artificial media structured on subwavelength scales and exhibiting electromagnetic properties sometimes contrasting with those of the conventional materials~\cite{Veselago,Elef,Caloz,Marques,Capolino}. An important class of such structures is presented by wire metamaterials based on regular two-dimensional (2D) or three-dimensional (3D) arrays of metallic wires~\cite{Pendry1998,Belov2003,Simovski2012}. In the simplest approximation, the electromagnetic properties of such materials are captured by the Drude model with the frequency-dependent permittivity~\cite{Pendry1998,Belov2002}
\begin{equation}\label{eq:Drude1}
    \varepsilon(\nu)=1-\frac{\nu_p^2}{\nu^2-j\nu\,\Gamma}\:,
\end{equation}
where $\Gamma$ measures Ohmic losses which are typically small at microwave frequencies, while $\nu_p$ is known as the {\it plasma frequency}. The propagation of waves at the frequencies below $\nu_p$ is strongly suppressed.

The response of wire metamaterials resembles that of an electron plasma in metals. In the latter case, however, $\nu_p$ is determined by the carrier density and cannot be modified. In contrast, the plasma frequency of a wire metamaterial is defined by the lattice period and radius of the wires, which makes it possible to flexibly tune the plasma frequency of such structures.

\begin{figure}[b]
	\center{\includegraphics[width=0.98\columnwidth]{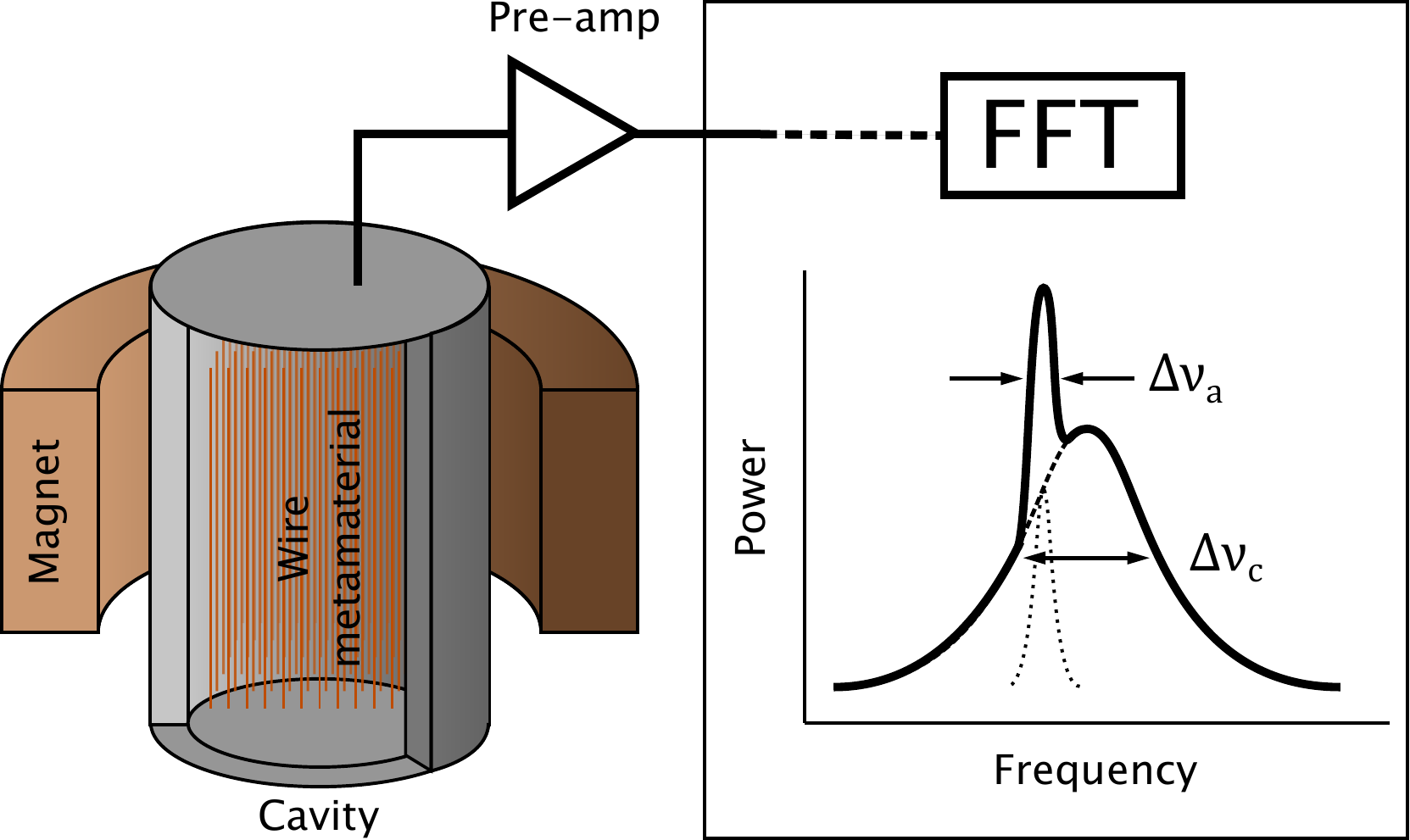}}
	\caption{Schematic of the microwave cavity dark matter axion search. The cavity bandpass, $\Delta \nu _c$ is determined by the quality factor of the cavity $Q$. The axion signal is broadened due to the virial velocity of dark matter in the Milky Way halo, $\Delta \nu_a / \nu_a \sim \beta _{vir}^2 \sim 10^{-6}$ . The frequency of the cavity is tuned in small steps, and the power spectrum at each central frequency, calculated by a Fast Fourier Transform, is integrated for sufficient time to achieve the desired signal-to-noise ratio for a particular axion-photon coupling and assumed local density.}
	\label{halo scheme}
\end{figure}

Recently, wire metamaterials have emerged in the context of ongoing searches of a hypothetical particle called {\it axion}.  This particle is a well-motivated candidate to constitute the dark matter of the universe~\cite{Feng2010} whose local density in our galactic halo is estimated to be $\rho_a \sim$ 0.45 GeV/cm$^3$. Axions are extraordinarily weakly coupled to matter and photons, and thus defy detection in the conventional reactor- or accelerator-based searches. They may however be detected by their resonant conversion to a weak quasi-monochromatic radio signal in a high-$Q$ microwave cavity permeated by a strong magnetic field (Fig.~\ref{halo scheme}), the resonant condition being that the cavity frequency equals the axion mass~\cite{Sikivie83}: $h\nu = m_a\,c^2$. For scale, 1 GHz = 4.136 $\mu$eV.

The axion mass is poorly constrained ranging from $10^{-9}$ to $10^{-3}$ eV~\cite{Irastorza2018} and, depending on the anticipated mass, different detection approaches have to be used. In the microwave cavity dark matter axion search \cite{Sikivie83}, the axion-photon conversion power depends on both unknown physics parameters like the axion mass $m_a$, and experimental parameters within the control of the operator:
\begin{equation}\label{eq:Axycoupling}
    P_{a\gamma} \propto (g_\gamma^2\rho_a m_a)(B^2VCQ)\:,
\end{equation}
where $g_\gamma$ is the dimensionless axion-photon coupling constant, of the order of unity, $B$ is the magnetic field, and $V$ the cavity volume. The cavity form factor $C$ represents the normalized squared overlap integral of the external magnetic field and the electric field of the cavity mode, a number between 0 and 1. For all current experiments, the signal optimistically will be of the order $10^{-23}$ Watts, putting a premium on maximizing all factors within the experimentalist's control, and exploiting state-of-the-art quantum limited or even sub-quantum limited receiver technology to maximize the signal-to-noise ratio~\cite{Backes21}. A detailed treatment of the microwave cavity technique, a description of current experiments and their limits has appeared in a recent review~\cite{Simanovskaia22}.

Previous and current experiments largely cluster in a little more than a decade of frequency, roughly between 0.5 and 5~GHz, which can be easily understood qualitatively. As the frequency of a cavity mode is inverse to its physical size, the lower limit on frequency is imposed by the diameter of the magnet bore in which the cavity resides. Regarding the practical upper limit of frequency, one can in principle use a smaller, higher frequency cavity in a large magnet, but since the volume of a cavity goes down as the inverse third power of its linear dimension, single small cavities incur a steep penalty in the conversion power. Arrays of multiple phased cavities and segmented cavities to maximize the utilization of magnetic volume have been attempted but are challenging from an engineering standpoint. There are other strategies as well to make large volume resonators~\cite{Kinon01,Jeong20} for higher frequencies [15,16], but none have been developed for operation above 10 GHz.

Recently, Lawson {\it et al.}~\cite{Lawson19} have proposed a novel solution to the conundrum of how to design a resonator that can be both sufficiently large to produce a detectable signal power, but for much higher frequencies than have been previously probed. In this scheme, the microwave cavity is replaced by a metamaterial consisting of a two-dimensional lattice of wires or equivalently, thin rods. Whereas in a conventional microwave cavity, the frequency of a mode is determined by the cavity’s size, the metamaterial is characterized by a plasma frequency, which is determined by its unit cell, a bulk property. In actual experiments, the metamaterial would reside in a metal enclosure, both to shield it from external radiofrequency noise and to maximize its quality factor $Q$. This appears to be a promising approach to cover the range of frequencies between 10 and 45 GHz, corresponding to the mass range of the axion of 40 to 180 ~$\mu$eV, predicted in one calculation to provide the dark matter density of the universe in a post-inflation scenario~\cite{Millar23, Buschmann2022}.

The actual feasibility of wire array metamaterials for the haloscope application is primarily defined by two factors. First, the quality factor of the cavity should be large enough to provide a sufficiently strong signal. First experimental studies have measured the quality factors around $Q\approx 220$~\cite{Balafendiev22} for a wire radius of 25~$\mu$m.  However, wires with a radius around 3~mm are anticipated to provide much higher quality factors $Q\sim 4000$ in the 10~GHz range~\cite{Millar23}, for which the projected quality factor at cryogenic temperatures can exceed $10^4$~\cite{Millar23}.

Another crucial ingredient for the axion search is the ability to tune the array in a practical manner over a useful dynamic range in frequency, which should be at least 10\%. The purpose of this study is therefore to investigate the pathways to flexibly tune the plasma frequency of the wire metamaterial. As a promising strategy, we consider a metamaterial composed of two interpenetrating wire sublattices which can be shifted relative to each other. Unlike the more intuitive tuning scheme involving changing the lattice period \cite{Ivzhenko2016}, the proposed method keeps the volume of the structure practically constant. As we prove numerically and experimentally, the proposed strategy allows us to tune the plasma frequency of the structure up to 16\% relative to the maximum frequency.

The rest of the Article is organized as follows. In Sec.~\ref{TheoryAndSimulation} we develop an analytical theory of metamaterials consisting of two interpenetrating wire lattices and confirm our predictions by the full-wave numerical simulations. Section~\ref{ExperimentalDetails} describes our experimental results showing good agreement with our analytical predictions. Finally, we conclude the paper with the discussion of our results and an outlook.

\section{Analytical and numerical studies}\label{TheoryAndSimulation}

\begin{figure}[ht]
	\center{\includegraphics[width=0.65\columnwidth]{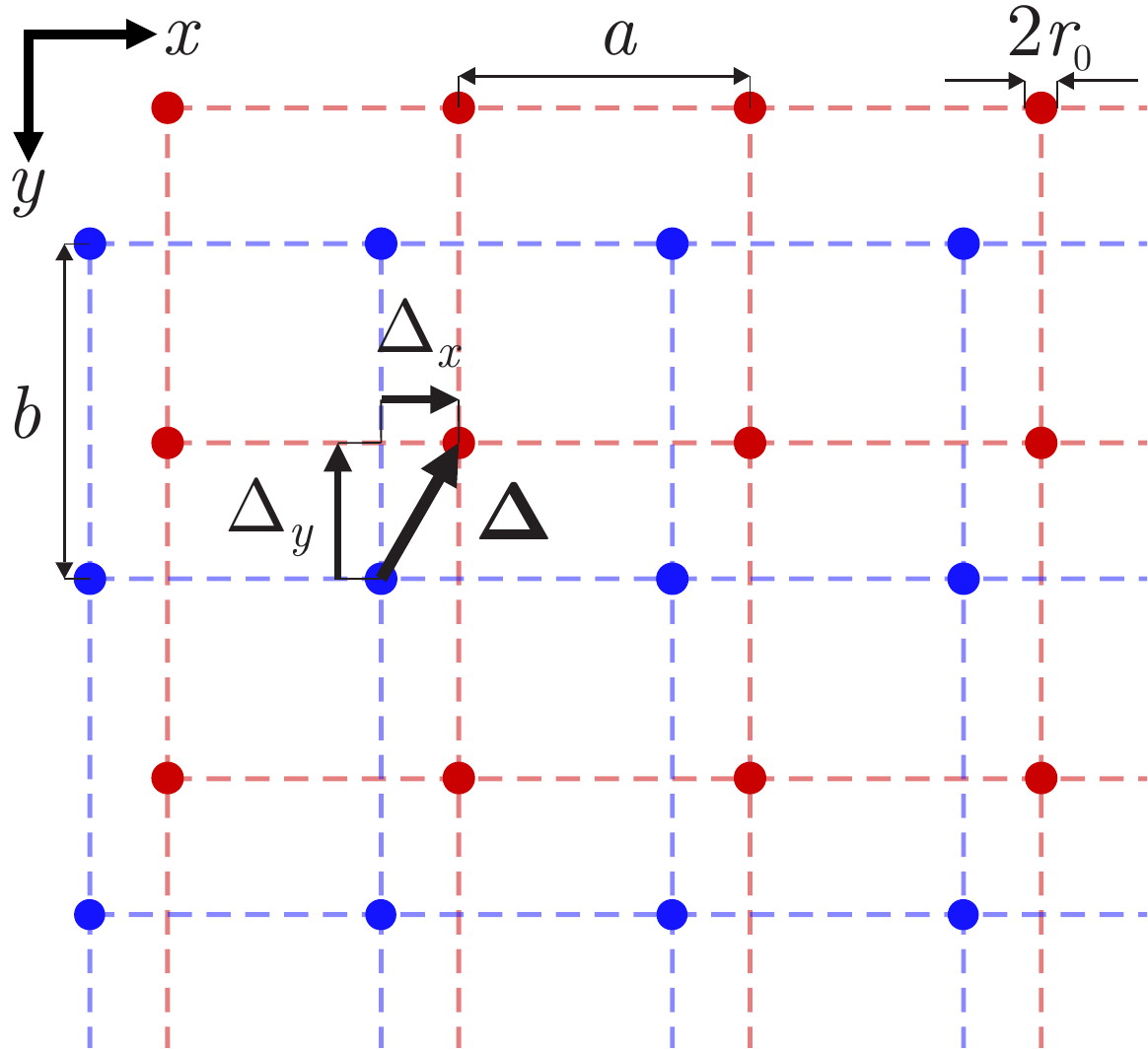}}
	\caption{Schematic of a wire medium consisting of two interpenetrating rectangular lattices with the unit cell sizes $a\times b$ shifted with respect to each other by the vector ${\bf \Delta}=(\Delta_x,\Delta_y)$. The wire radius $r_0$ is assumed to be significantly smaller than the periods $a$ and $b$.}
	\label{fig:scheme}
\end{figure}

Previously, the analytical description was developed for the wire medium based on the rectangular lattice of thin wires~\cite{Belov2002}. Below, we generalize this theory for the structure consisting of two interpenetrating wire lattices as depicted in Fig.~\ref{fig:scheme}. Each of sublattices has $a\times b$ unit cell shifted with respect to each other by the offset vector ${\bf \Delta}=(\Delta_x,\Delta_y)$. Our goal is to calculate the plasma frequency of this metamaterial as a function of the lattice geometry governed by the parameters $a$, $b$, $\Delta_x$ and $\Delta_y$.

In our analysis, we assume that the wires are thin, i.e. $r_0\ll a$ and $r_0\ll b$. Therefore, we assume that the current distribution in a wire does not depend on the azimuthal angle which corresponds to the lowest-frequency current mode. A time-varying current $I\,e^{-jq_z\,z+j\om\,t}$ in a single wire creates a parallel AC field given by
\begin{equation}\label{eq:Efield}
    E_z=-\frac{\eta\,\kap^2}{4\,k}\,H_0^{(2)}(\kap\,R)\,I\,e^{-jq_z\,z+j\om\,t}\:,
\end{equation}
where the SI system of units and the $e^{j\om\,t}$ time convention are adopted, $k=\omega/c$, $q_z$ is the wave number along the axis of the wire,  $\kap=\sqrt{k^2-q_z^2}$, $R = \sqrt{x^2+y^2}$ and $H_0^{(2)}$ is the Hankel function of the second kind.

The electric field produced by the wire couples to the currents in the other wires giving rise to the collective mode. If the structure is periodic, the amplitudes of the currents obey Bloch's theorem:
\begin{equation}\label{eq:Bloch}
    \begin{pmatrix}
    I^{(A)}_{mn}\\ 
    I^{(B)}_{mn}
    \end{pmatrix}
    =
    \begin{pmatrix}
    I^{(A)}\\
    I^{(B)}
    \end{pmatrix}
    e^{-j{\bf q}\cdot{\bf R}_{mn}}\:,
\end{equation}
where $I^{(A)}$ and $I^{(B)}$ stand for the current amplitudes in the two sublattices.

On the other hand, since the wires are perfectly conducting, boundary conditions require that the total electric field at the surface of the wire vanishes. Writing the total field explicitly and dividing the result by $-\eta\,j\,\kap^2/(2k)$, we recover:
\begin{gather}
    C_{{\rm rec}}\,I^{(A)}+C({\bf \Delta})\,I^{(B)}=0\:,\label{eq:Int1}\\
    C_{{\rm rec}}\,I^{(B)}+C(-{\bf \Delta})\,I^{(A)}=0\:,\label{eq:Int2}
\end{gather}
where 
$C_{{\rm rec}}$ captures the contribution from the wires of the same sublattice as the wire under consideration at its surface, including its own field, and $C({\bf \Delta})$ is responsible for the wire interaction with another sublattice.

Setting the determinant of the system Eqs.~\eqref{eq:Int1}, \eqref{eq:Int2} to zero, we derive the dispersion equation:
\begin{equation}\label{eq:Dispersion}
 C_{{\rm rec}}\pm \sqrt{C({\bf \Delta})\,C(-{\bf \Delta})}=0\:.
\end{equation}

Note that the $+$ or $-$ sign choice in this equation defines the relative phase of the currents $I^{(A)}$ and $I^{(B)}$ in the two sublattices. 

Here, the interaction constant for the rectangular lattice $C_{{\rm rec}}$ can be written as~\cite{Belov2002}:
\begin{multline}\label{eq:Rect}
  C_{{\rm rec}}=\frac{1}{\pi}\,\ln\left(\frac{b}{2\pi\,r_0}\right)+\frac{1}{k_x^{(0)}\,b}\,\frac{\sin \left(k_x^{(0)}\,a\right)}{\cos \left(k_x^{(0)}\,a\right)-\cos \left(q_x\,a\right)}\\
  +\sum\limits_{n\not=0}\,\left\lbrace\frac{1}{k_x^{(n)}\,b}\,\frac{\sin \left(k_x^{(n)}\,a\right)}{\cos \left(k_x^{(n)}\,a\right)-\cos \left(q_x\,a\right)}-\frac{1}{2\pi\,|n|}\right\rbrace\:,
\end{multline}
where $k_x^{(n)}=-j\,\sqrt{\left(q_y^{(n)}\right)^2-\kap^2}$ and $q_y^{(n)}=2\pi\,n/b+q_y$.

Using the Poisson summation formula we also derive the expression for the interaction constant of the two lattices:
\begin{multline}
\label{eq:Intnew}
    C({\bf \Delta})=\\
    -\sum\limits_{n=-\infty}^{\infty}\frac{e^{jq_y^{(n)}\,\Delta_y}}{k_x^{(n)}\,b}\,\frac{e^{jq_x a}\,\sin \left(k_x^{(n)}\Delta_x\right)+\sin\left[k_x^{(n)}\,\left(a-\Delta_x\right)\right]}{\cos \left(q_x\,a\right)-\cos \left(k_x^{(n)} a\right)}\:.
\end{multline}

\begin{figure*}[t]
\includegraphics[width=\textwidth]{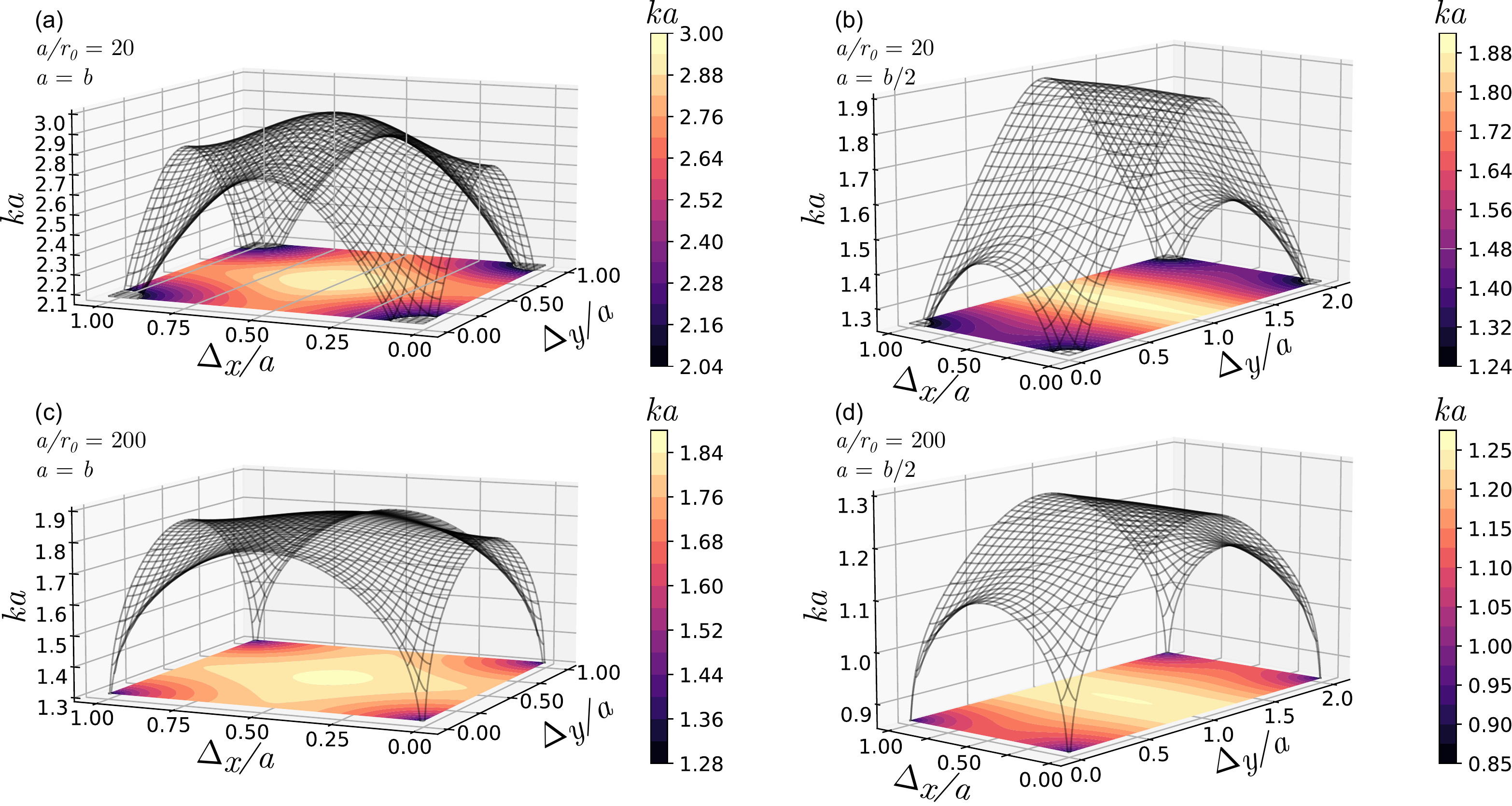} 
 \caption{
Results of the semi-analytical Matlab calculations for four infinite lattices of wires. 
 Contours and wiremesh surface show the normalised plasma frequency $k a$ as a function of the relative sublattice displacement $\Delta_x$ and $\Delta_y$.
 Subfigures (a) and (b) show the results for a medium with $a/r_0 = 20$, close to the point where wires can no longer be considered thin. Subfigures (c) and (d) depict the change of the plasma frequency for a medium with $a/r_0 = 200$, well within the bound of our theory being applicable.
 Sublattices on the two subfigures to the left are square, with periods $b=a$. Sublattices to the right are elongated, so that $b=2a$. 
 As can be seen, for square sublattices movement along any trajectory leads to a noticeable frequency shift. For rectangular sublattice, however, frequency tuning mainly corresponds to movement along $y$, with movement along $x$ bringing little change in frequency (see Table \ref{tab:percentage_an} below).}
 
     \label{fig:aspect} 
 \end{figure*}

The dispersion equation Eq.~\eqref{eq:Dispersion} together with the expressions for the interaction constants Eqs.~\eqref{eq:Rect}, \eqref{eq:Intnew} fully determines the dispersion properties of the wire medium sketched in Fig.~\ref{fig:scheme}. However, the key parameter relevant for plasma haloscopes is the plasma frequency $\nu_p$. It is obtained as the lowest-frequency root of the dispersion equation for the zero wave vector ${\bf q}=0$. In such case, since the wires in the two sublattices are identical, $C(-{\bf \Delta})=C({\bf \Delta})$ which yields
\begin{equation}\label{eq:QZero}
    C_{{\rm rec}}\pm C({\bf \Delta})=0\:.
\end{equation}
Using the identity
\begin{equation*}
    \frac{\sin\alpha}{\cos\alpha-1}=-\cot\frac{\alpha}{2}\:,
\end{equation*}
we simplify the dispersion equation further and obtain:
%
\begin{multline}
\label{eq:PlasmaFrq}
    \frac{1}{\pi}\,\ln{\left(\frac{b}{2\pi\,r_0}\right)}-\frac{1}{k_x^{(0)}\,b}\,\cot\left(\frac{k_x^{(0)}\,a}{2}\right)\\
    -\sum\limits_{n\not=0}\,\left[\frac{1}{k_x^{(n)}\,b}\,\cot\left(\frac{k_x^{(n)}\,a}{2}\right)+\frac{1}{2\pi\,|n|}\right]\\
    \mp \sum\limits_{n=-\infty}^{\infty}\,\frac{\exp\left(2\pi j n\Delta_y/b\right)}{k_x^{(n)}\,b}\,\frac{\cos\left[k_x^{(n)}(a/2-\Delta_x)\right]}{\sin\left(k_x^{(n)}\,a/2\right)}=0\:.
\end{multline}
%
Here, $k_x^{(n)}=-j\sqrt{\left(\frac{2\pi n}{b}\right)^2-k^2}$, and the unknown parameter is the normalized frequency $k=\omega/c$. The plasma frequency is calculated from the lowest-frequency root $k_{\text{min}}$ of this equation as $\nu_{pl}=ck_{\text{min}}/(2\pi)$.

The lowest-frequency root of this equation yields the plasma frequency.

At this point, we comment on the sign choice in Eq.~\eqref{eq:PlasmaFrq}. A straightforward approach is to solve Eq.~\eqref{eq:PlasmaFrq} with respect to $k$ numerically for both possible signs and then choose the smallest root. However, the sign can  also be chosen based on physical reasoning. The lowest-frequency mode corresponding to the plasma frequency is characterized by in-phase currents in the two sublattices. Hence, the dispersion equation should be $C_{{\rm rec}}+C({\bf \Delta})=0$ which corresponds to the upper sign choice in Eqs.~\eqref{eq:QZero}. Hence, the correct sign in Eq.~\eqref{eq:PlasmaFrq} is the upper one, i.e. minus.


\begin{table}[ht!]
\begin{tabular}{ |c||c|c|c|  }
 \hline
 {\makecell{\\Lattice \\parameters}}{\makecell{Change of\\ ($\Delta_x$,$\Delta_y$)}}
 & \makecell{($a/2,b/2$) \\to (0,$2r_0$) \\
 \includegraphics[width=0.12\columnwidth]{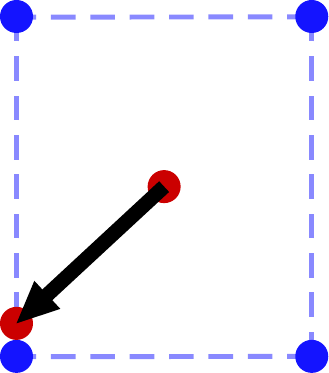}}
 &\makecell{($a/2,b/2$) \\to (0$,b/2$)\\
 \includegraphics[width=0.12\columnwidth]{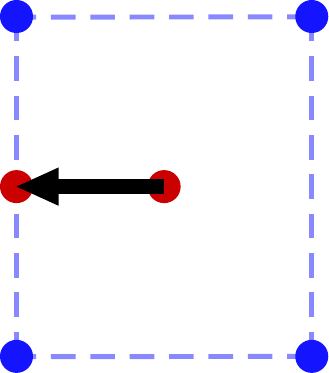}}
 &\makecell{($a/2,b/2$) \\to ($a/2$,0)\\
 \includegraphics[width=0.12\columnwidth]{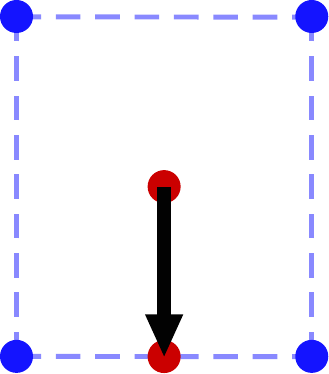}} \\
 \hline
 $a/r_0 = 20$, $b = a$   & 29.2\% & 7.5\% & 7.5\%\\
 $a/r_0 = 20$, $b = 2a$  & 33.7\% & 0.3\% & 22.5\%\\
 $a/r_0 = 200$, $b = a$  & 29.3\% & 2.7\% & 2.7\%\\
 $a/r_0 = 200$, $b = 2a$ & 31.0\% & 0.1\% & 10.9\%\\
 \hline
\end{tabular}
 \caption{Tuning percentage of the plasma frequency in the semi-analytical Matlab calculations, defined as $(\nu_{\text{max}}-\nu_{\text{min}})/\nu_{\text{max}}$ for various cases illustrated in Fig.~\ref{fig:aspect}. The first column shows the tuning achievable by bringing the two sublattices in direct contact. The second and third columns show tuning by moving the second sublattice from the maximum frequency configuration to the edge of the unit cell along $x$ and $y$ respectively.}
 \label{tab:percentage_an}
\end{table}

The result for the rectangular lattice~\cite{Belov2002} is recovered from Eq.~\eqref{eq:PlasmaFrq} by omitting the last two terms characterizing the interaction with another sublattice.


To explore the dependence of the plasma frequency on the parameters of the wire medium, the resulting equation has been solved numerically using Matlab software package for several values of $r_0/a$ and $a/b$. Figure~\ref{fig:aspect} shows how the plasma frequency of an infinite medium changes with the relative shift of the two sublattices. Subfigures \ref{fig:aspect}a and \ref{fig:aspect}b show the change in the frequency for the case where the ratio of unit cell period to the wire radius $a/r_0 = 20$. For subfigures \ref{fig:aspect}c and \ref{fig:aspect}d, the wires are thinner: $a/r_0 = 200$. As can be seen, the cases with a rectangular sublattice unit cell ($b = 2a$) produce a dramatic change in frequency when tuned along the wider side and nearly no change when tuned along the shorter one (Table \ref{tab:percentage_an}). The comparison of the two radii also shows that the relatively thicker wires enable greater tuning when sublattices are moved along the larger period $b$ (i.e. in the $y$ direction) from the position of the maximum plasma frequency. In order to verify the obtained results, the same unit cells were modeled using COMSOL Mutliphysics Eigenmode Solver, with the resulting discrepancy of the semi-analytic results with the numerical ones being less than 0.2\%.


  

\section{Experimental Description}\label{ExperimentalDetails}

\begin{figure}[h]
	\center{\includegraphics[width=0.98\columnwidth]{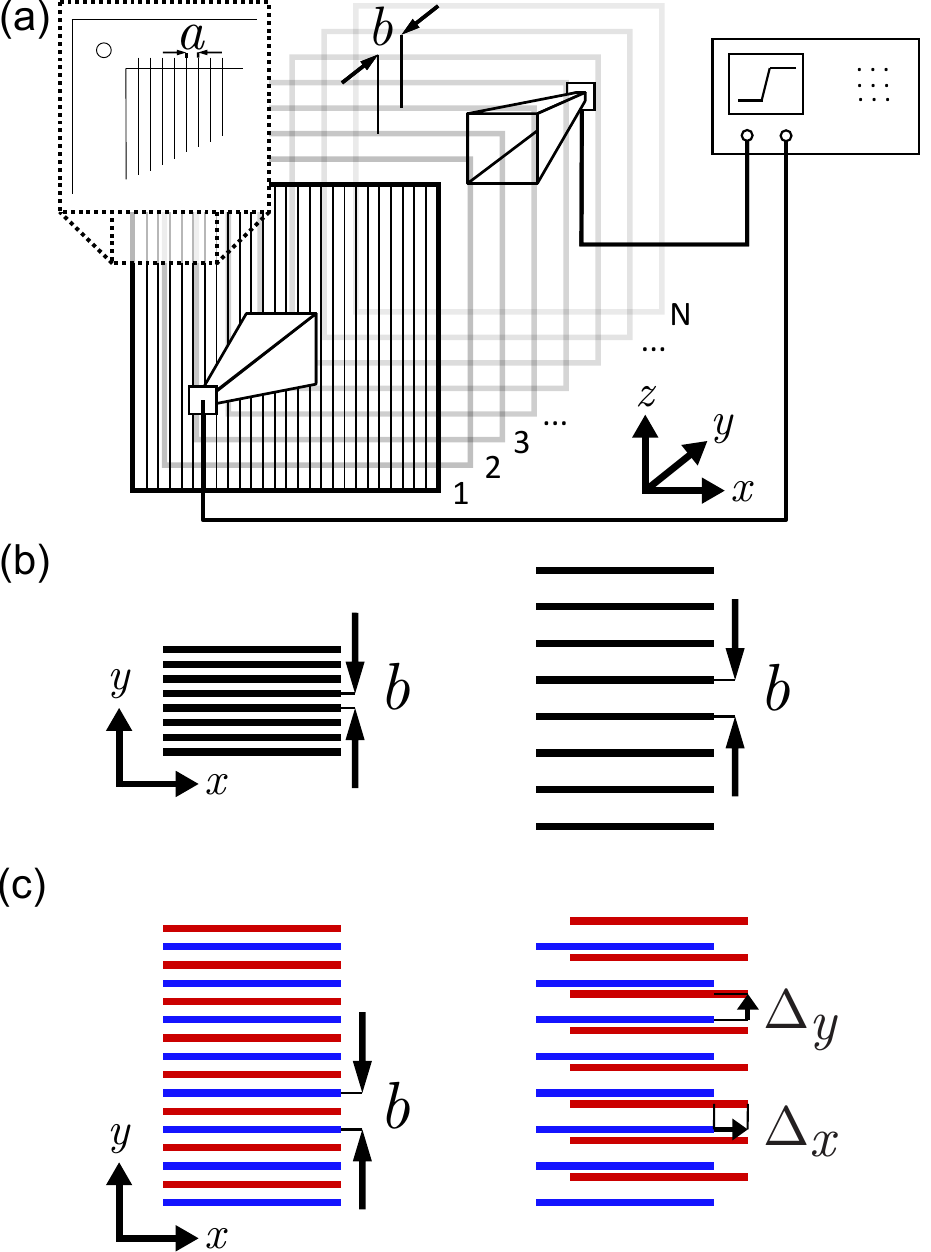}}
	\caption{
     a) The geometry for the measurements of $S _{21}$ for an array of N planes. The inset on the top-left shows the construction of a single wire plane. The wires are supported by the metallic wire frame but are not in electrical contact with it. b) The first set of measurements, looking at the change in plasma frequency as a function of spacing between planes of wires $b$. c) The second set of measurements, done with two sets of planes. The change is plasma frequency is measured as a function of relative plane sublattice offset ${\bf \Delta}$=$(\Delta_x,\Delta_y)$.}
	\label{fig:meta scheme}
\end{figure}

The experiments mapping out the dependence of the plasma frequency on the unit cell were carried out with the microwave setup depicted in Fig.~\ref{fig:meta scheme}. The three-dimensional wire array was built up by stacking planes of regularly spaced wires; this allowed the lattice to be conveniently reconfigured by changing spacers between planes, shifting alternate planes, etc.

\begin{figure}[ht]
	\center{\includegraphics[width=\columnwidth]{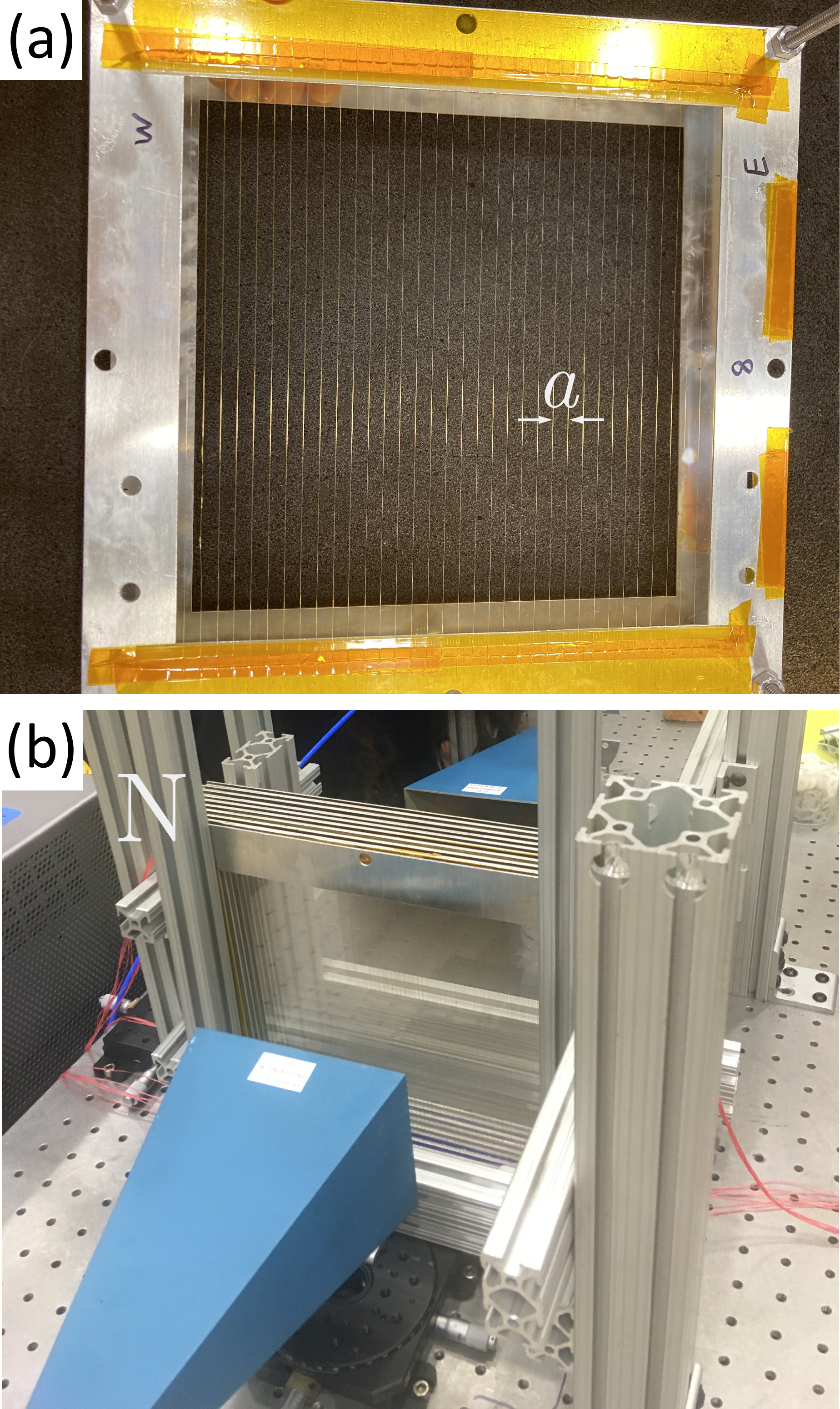}}
	\caption{(a) Metal frame being strung with the 50~$\mu$m gold-on-tungsten wires placed with a period  $a$~=~5.88~mm. (b) Setup for the $S_{21}$ measurements of N planes  in the close-pack configuration, i.e. without spacers.}
	\label{fig:setup}
\end{figure}

The wire planes, 20 in total, were constructed by fabricating square aluminum frames of 203 (254)~mm inner (outer) edge length and 1.23 mm~thickness. The frames were then strung with gold-on-tungsten wires \cite{Lumametal} of radius and spacing in the plane of ($r_0$ , $a$) = (25~$\mu$m, 5.88~mm). The wires were glued onto a thin plastic bridge on either side of the frame, from which they were thus electrically isolated, see Fig. ~\ref{fig:setup}(a).

The metamaterial parameters are extracted from the measured transmission spectra, i.e. the $S_{12}$, one of the four scattering parameters for a two-port device. Specifically this is the ratio of the output of the second port relative to the input of the first port, $ V_2^+ / V_1^+$, and is conveniently expressed as the scalar logarithmic gain, $g = 20·log _{10} (|S_{21} |)$~[dB]. Within a Drude model, the complex frequency dependent permittivity is 
given by $\epsilon(\nu) = \epsilon'(\nu)-j\epsilon''(\nu)$, where

\begin{equation}\label{eq:Drude}
     \epsilon'(\nu)= 1-\frac{\nu_p^2}{\nu^2+\Gamma^2} \mbox{ and } \epsilon''(\nu) = (\frac{\nu_p}{\nu})^2\frac{\frac{\Gamma}{\nu}}{1+(\frac{\Gamma}{\nu})^2}.
\end{equation}

The $S_{12}$ measurements were performed with a vector network analyzer \cite{VNA}, and matched waveguide horn antennas \cite{Horns}. All measurements were performed in normal incidence geometry, i.e. beam propagation perpendicular to the wire array. As the waveguide horn antennas have a beam spread of $\pm 8^\circ$ at 3~dB, tests were performed to determine whether scattering from the inside edge of the frames was contributing to the spectra. This was done by comparing the fits with and without a rectangular collimater of microwave absorbing material of 7.5~cm (horizontal) $\times$ 5.0~cm (vertical). No evidence was seen for scattering from the frames, and thus all measurements were performed without the collimator in place. Before and after each measurement with the wire array in place, the baseline transmission was recorded with the wire array removed; the two were averaged and then subtracted from the spectrum measured with the wire array in place to yield the $S_{12}$ to be fit.

\begin{figure}[ht]
	\center{\includegraphics[width=\columnwidth]{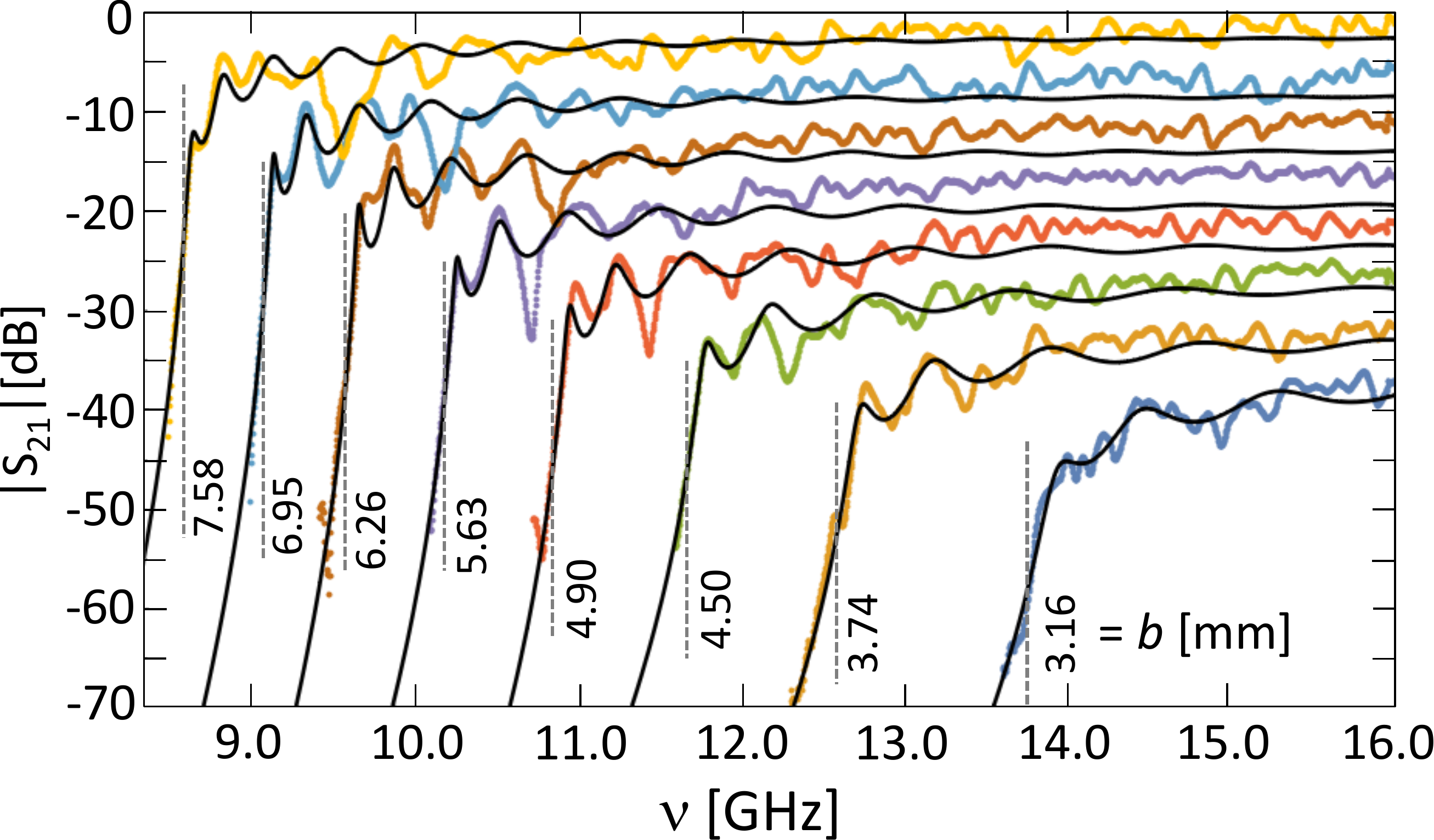}}
	\caption{The $S_{12}$ spectra for a rectangular lattice, where the wire spacing in the plane $a$ is fixed, and the spacing between planes $b$ is varied. The experimental data is represented in color; the fitted calculation in black. The vertical dotted lines indicate the fitted plasma frequency for each spectrum. The spectrum for $b$ = 7.58~mm corresponds exactly to the vertical scale; each subsequent spectrum is offset by -5~dB for clarity.}
	\label{fig:S12}
\end{figure}

The measured $S_{12}$ spectrum is fit with the calculated transmission through a uniform dielectric of complex permittivity $\epsilon'$ \cite{Yahalom10}, returning values of $\nu_p$, $\Gamma$ and $d$, with errors; see Figure \ref{fig:S12} for representative spectra and their fits. The sharp transition edge of the spectrum most sensitively encodes the plama frequency $\nu_p$ and the loss term $\Gamma$ ; the effective width of the array $d$ by the first few oscillations above the cutoff.

Extensive measurements were performed on two different methods of modifying the unit cell. The first (Fig. \ref{fig:meta scheme}b) restricted the unit cell to be rectangular, maintaining the same wire spacing $a$ in the plane, but changing the spacing between the planes, $b$. The second  (Fig. \ref{fig:meta scheme}c) involved ganging alternate planes together, and then translating the two groups relative to one another by an offset ($\Delta_x$, $\Delta_y$), i.e. a parallel and perpendicular shift. Each set of measurements will be discussed in turn.

\begin{figure}[ht]
	\center{\includegraphics[width=1.0\columnwidth]{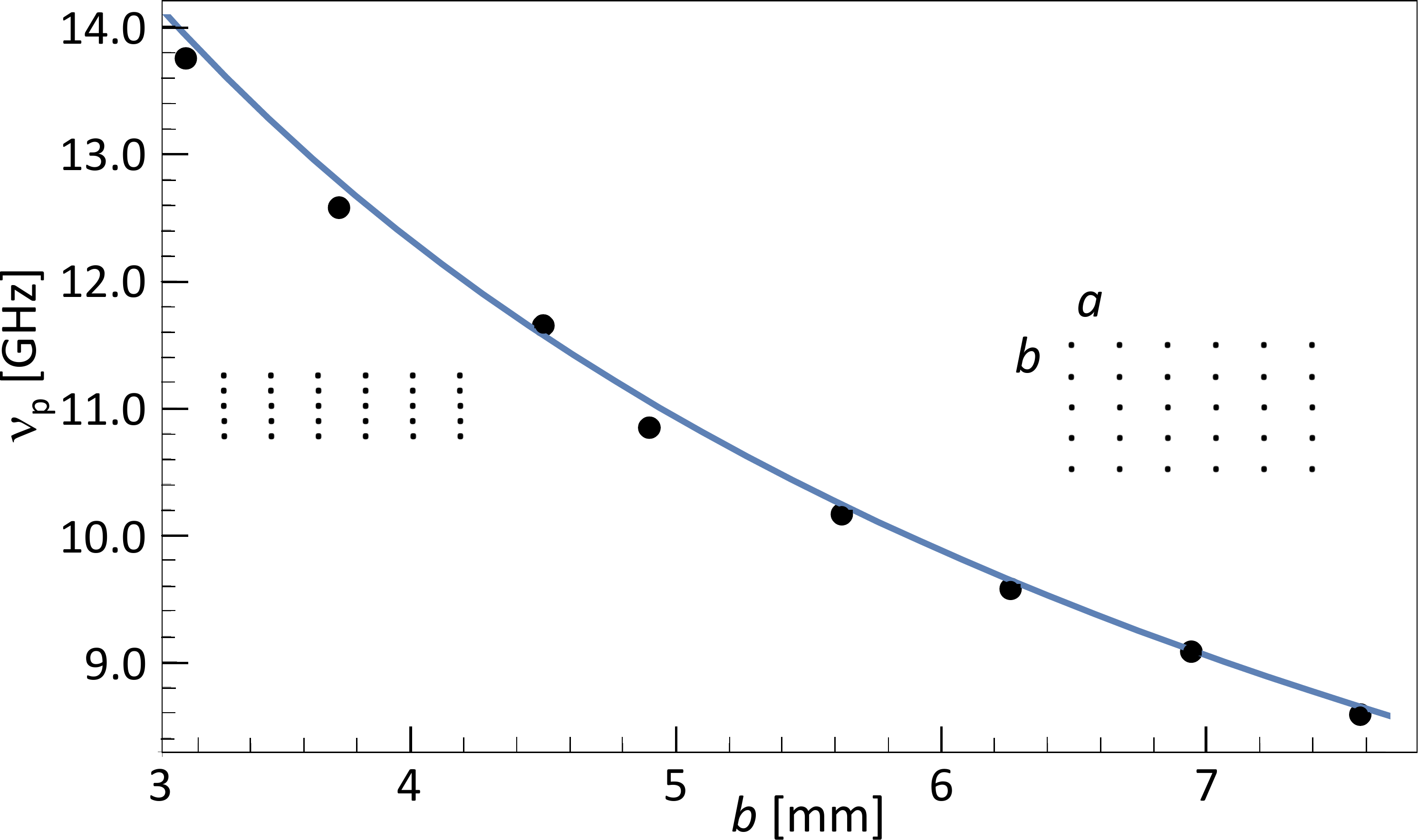}}
	\caption{Plasma frequency for a rectangular unit cell with $a$ = 5.88 mm, and $b$ variable. Ideograms indicate larger and smaller $b$ are associated with lower and higher plasma frequencies respectively. Solid dots represent data; the error bars are smaller than the symbols. Blue line represents the absolute prediction of the semianalytic theory \cite{Belov2003}.}
	\label{fig:vpvsb}
\end{figure}

(i) Varying the interplane spacing. Measurements were made with 20 wire frames, for 8 different values of the interplane spacing $b$, varying from 3.16 to 7.58~mm. The waveguide horn antennas were placed 48 cm apart for all measurements. The plasma frequency as a function of $b$ is shown in Figure \ref{fig:vpvsb}. The blue curve represents the prediction of a semianalytic theory \cite{Belov2003}. Note that this is an absolute prediction, with no adjustable parameters; the agreement with data is at the 1\% level.

While uniformly changing the spacing between planes can produce a large dynamic range in frequency, in this case a 60\% increase in frequency by decreasing the spacing from 7.6~mm to 3.2~mm, this scheme presents difficulties from the perspective of designing a practical axion haloscope. Tuning the haloscope in this manner does not conserve volume of the active metamaterial resonator; in this example the conversion volume is reduced by a factor of 2.4 from the lowest to highest frequency, 8.6-13.8~GHz. This represents a large loss of active volume and thus axion-plasmon conversion power, which is precisely what the plasma haloscope concept was designed to circumvent. Furthermore, it is difficult to envision a mechanical solution by which the hermetic microwave enclosure containing the array and which must closely conform to its boundary (roughly within a wire spacing) can be continuously expanded and contracted as the haloscope is tuned. For a full discussion of the implementation of a wire array metamaterial in an axion haloscope, and specifically the integrated design of the wire array within its cavity, see \cite{Millar23}.

(ii) Varying the position of alternate planes. The dependence of the plasma frequency with unit cell was also examined by uniformly translating alternate planes of the array in both parallel and perpendicular directions to the microwave beam. Figure \ref{fig:scheme}(a) depicts the coordinate ($\Delta_x$,~$\Delta_y$) system for the translation, where $\Delta_x$ denotes the relative translation of planes parallel to one another, keeping their separation fixed, and $\Delta_y$ denotes the relative translation in the microwave propagation direction and which thus changes the relative spacing of the planes. For ($\Delta_x$,~$\Delta_y$) = (0,~$b/2$), a rectangular lattice of wire spacing ($a$,~$b/2$) is recovered; by symmetry, one only needs to map out the unit cell from (0,~0) to ($a/2$,~$b/2$), but extending the measurements is a valuable check on possible systematic errors. The experimental setup is shown in Fig.~\ref{fig:setup}(b).

\begin{figure}[ht!]
	\center{\includegraphics[width=\columnwidth]{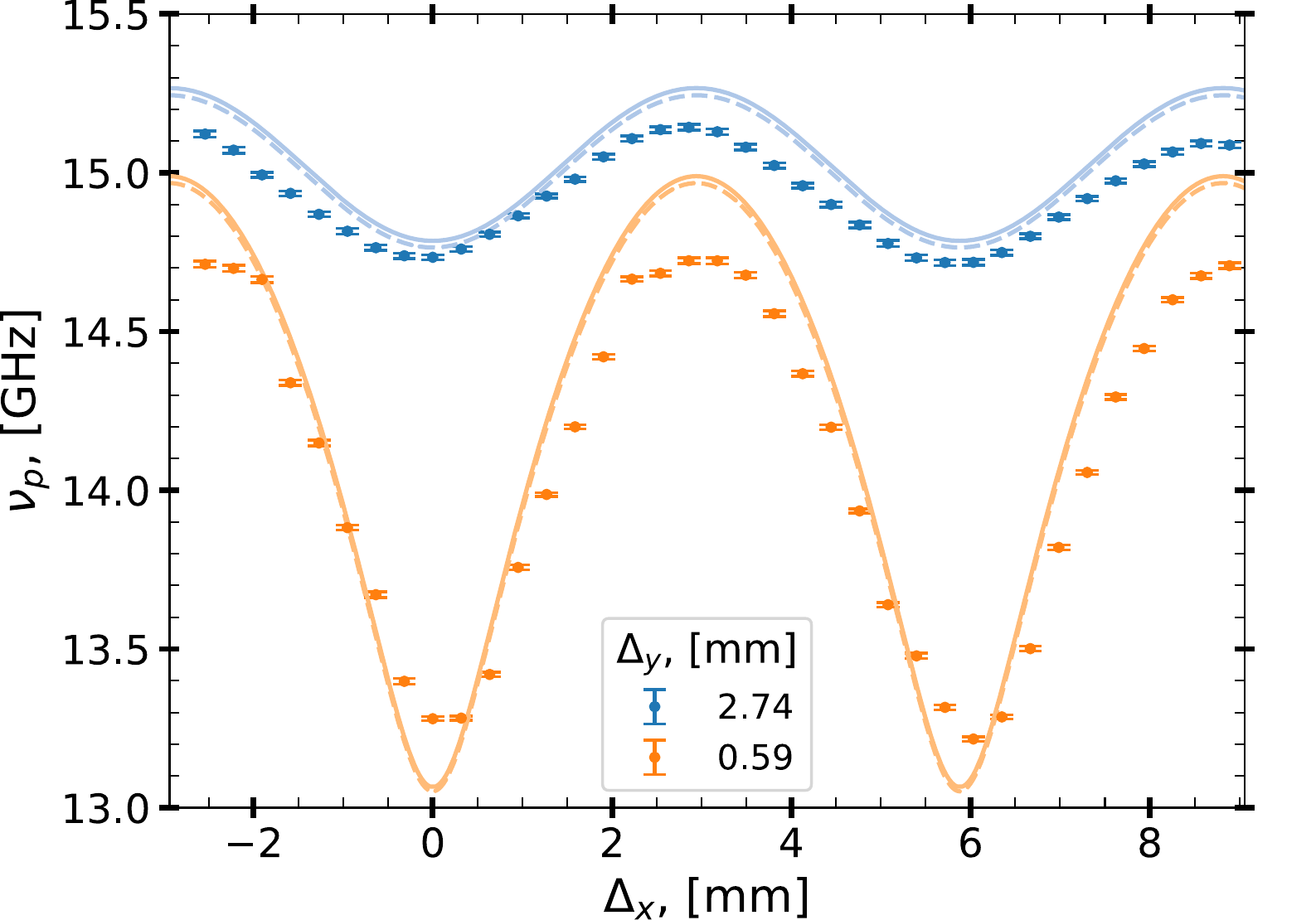}}
	\caption{Data and simulations for stacking the wire frames without spacers, ($a$, $b$) = (5.88~mm, 5.48~mm). Blue data points and curves pertain to the configuration where all the frames face in the same direction; orange data points and curves where alternate frames face in the opposite direction, bringing the wires into close proximity. The solid lines show semianalytical result calculated using Matlab, the dashed lines show numerical result calculated using Comsol. The discrepancy between the two is less than 0.15\% in both cases.}
	\label{fig:small_UC}
\end{figure} 

The plasma frequency $\nu_p$ was mapped out for two different lattices. Figure \ref{fig:small_UC} represents the lattice ($a$, $b$) = (5.88~mm, 5.48~mm) where the wire frames were stacked directly on top of one another, first with all the planes facing in the same direction ($\Delta_y$ = 2.74~mm), and then with alternate planes facing in the opposite direction ($\Delta_y$ = 0.59~mm). In the latter case, the actual planes of wires themselves come into near-contact with one another; the wires are separated only by a layer of Kapton tape on top of each bridge, and a thin polyethylene sheet to protect the wires from damage as the frames were shifted relative to one another from one measurement to another. After several thousand restacking and shifting operations over the course of the project, the wires were no longer entirely taut, and therefore the distribution of all wire positions relative to their fiducial location was measured on a microscope stage to determine their standard deviation along a single coordinate, $\sigma_y$ = 293 $\mu$m. This was incorporated as a correction to the mean distance for the case of alternate places facing one another ($\Delta_y$ = 0.59~mm), where the plasma frequency changes most rapidly as opposite pairs of wires approach one another. The correction is negligible at large wire separations.

\begin{figure}[h]
	\center{\includegraphics[width=\columnwidth]{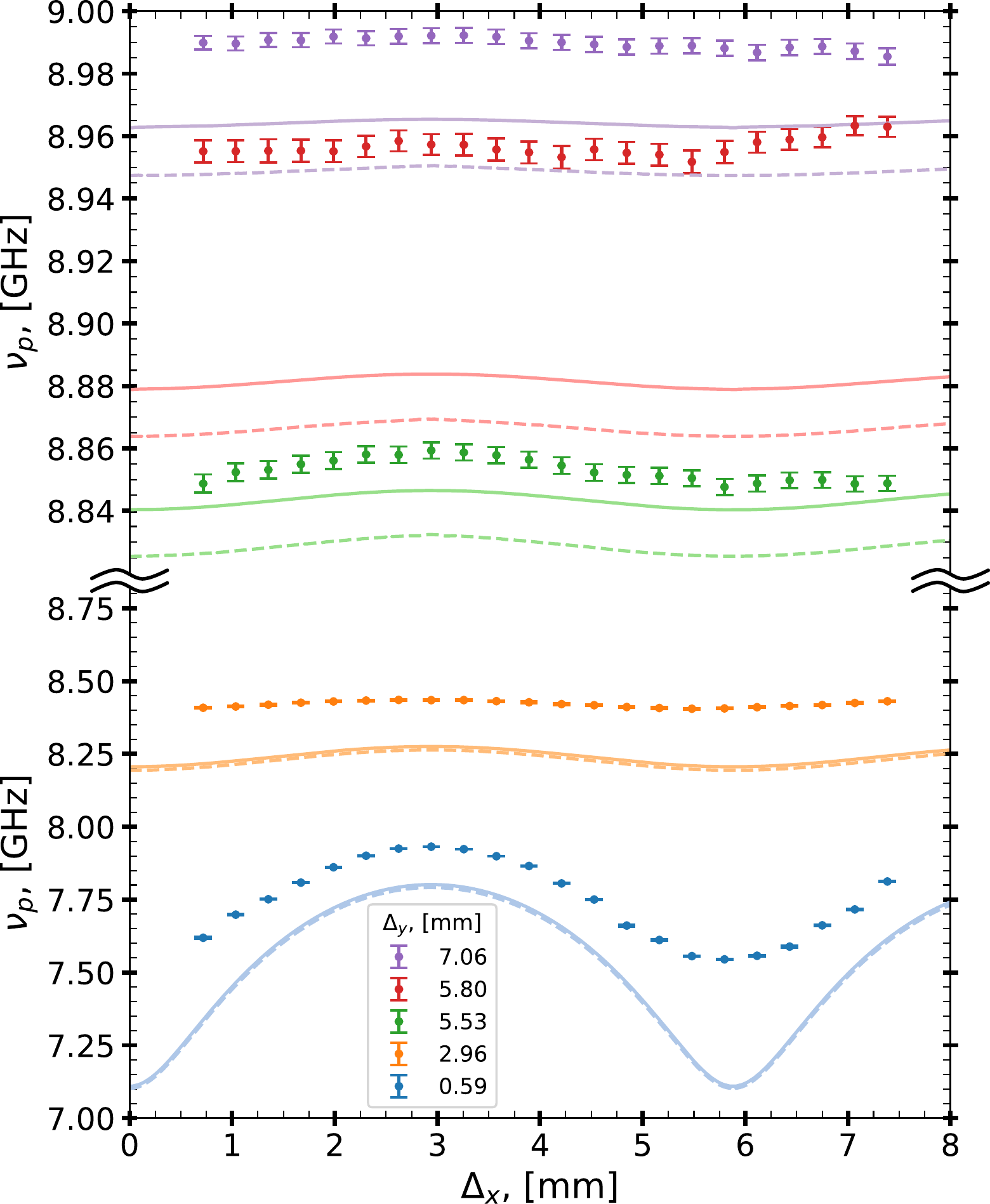}}
	\caption{Data and Matlab simulations for excursions from the unit cell ($a$, $b$) = (5.88~mm, 14.12~mm). The measurements and their corresponding predictions are in the same color. Minima in the plasma frequency correspond to alignment of wires in the $y$ direction. The solid lines show semianalytical result calculated using Matlab, the dashed lines show numerical result calculated using Comsol. The resulting discrepancy of the semianalytic results with the numerical ones is less than 0.2\% for all five cases. 
     }
	\label{fig:large_UC}
\end{figure} 

A second series of measurements was performed to map out the plasma frequencies for excursions from a larger rectangular lattice, ($a$, $b$) = (5.88~mm, 14.12~mm). The microwave horns were positioned 37 cm apart for this series. Here, alternate planes faced in opposite directions, and the spacing $\Delta_y$ was adjusted in discrete steps while keeping $b$ fixed, using a series of spacers. The transverse registration of the odd and even frames was determined by a micrometer-driven translation stage on either side of the array. Translation stages at the front and back of the array maintained the proper longitudinal dimension of the array for each measurement. This was necessary to ensure that the slight compressibility of the stack of frames and spacers did not introduce variations in $\Delta_y$ between successive measurements in $\Delta_x$, as the array needed to be relaxed longitudinally to shift alternate planes to their new position and then recompressed for the next measurement.

In contrast with tuning the metamaterial by varying the interplane spacing, modifying the unit cell by translating alternate planes preserves the volume of the array, and thus should be more amenable to implementation in an axion haloscope. (In fact, this method is not strictly volume-preserving but very nearly so; over the full range of unique unit cells, the array widens by $a$/2 in the x direction, and for an even number of planes, extends by $b$ in the y direction.) Relative to varying interplane spacing, in this case, dynamic range is sacrificed, but with the parameters explored here, the accessible frequency range would be of practical interest for the haloscope application.

The data and simulations for this configuration are shown in Figure \ref{fig:large_UC}. The various combinations of spacers enabled $\nu_p$ to be mapped out in fine steps in $\Delta_x$ for five values of $\Delta_y$, ranging from the wire planes placed close nearby ($\Delta_y$ = 0.59~mm) to equally spaced ($\Delta_y$ = $b/2$ = 7.06~mm). The overall agreement between data and simulations is very good, ranging from a few percent when wires are in close proximity ($\Delta_y \rightarrow 0$), to a few parts per mil at the largest spacings ($\Delta_y \rightarrow b/2$). It is also noteworthy that the measurement and fitting of the $S_{21}$ spectra in the latter case display the predicted oscillatory shape exactly, even for a magnitude of the oscillation as small as $10^{-3}$ in $\nu_p$, lending confidence to the experimental procedure and analysis. The discrepancies in the absolute plasma frequency at larger separations can be ascribed to limitations to how accurately the plane separations $\Delta_y$ can be measured; at the smallest separation $\Delta_y$ = 0.59~mm this is compounded by slackening of the wires after repeated handling and positioning operations, which is most clearly seen in the filling in of the minima of the curves.

\begin{table}
\begin{tabular}{ |c||c|c|c|c|  }
 \hline
 \makecell{Lattice \\parameters} 
 & \makecell{($a/2$,$b/2$) \\to (0,0.59)\\
 \includegraphics[width=0.12\columnwidth]{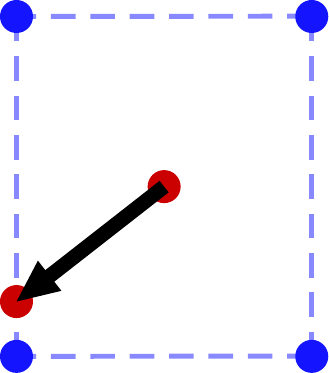}} 
 &\makecell{($a/2$,$b/2$) \\to (0,$b/2$)\\
 \includegraphics[width=0.12\columnwidth]{figures/dir2.pdf}}
 &\makecell{($a/2$,$b/2$) \\to ($a/2$,0.59)\\
 \includegraphics[width=0.12\columnwidth]{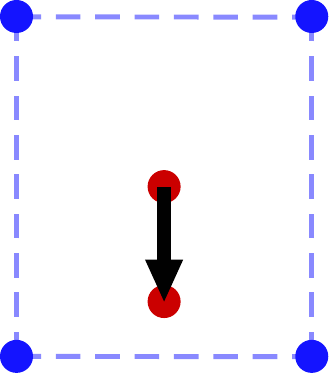}}
 &\makecell{(0,$b/2$) \\to (0,0.59)\\
 \includegraphics[width=0.12\columnwidth]{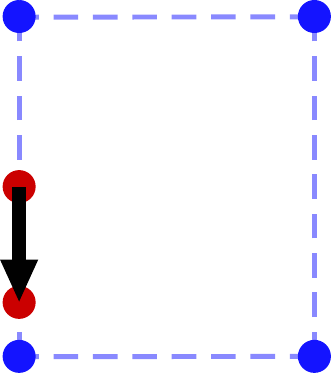}}\\
 \hline
  \makecell{($a$, $b$) =\\ (5.88~mm,\\ 5.48~mm)}   & 12.72\% & 2.81\% & 2.77\% & 9.87\%\\
   \hline
  \makecell{($a$, $b$) =\\ (5.88~mm,\\ 14.12~mm)}  & 16.09\% & 0.08\% & 11.79\% & 16.07\%\\

 \hline
\end{tabular}
 \caption{Tuning percentage, defined as $(\nu_{\text{max}}-\nu_{\text{min}})/\nu_{\text{max}}$ for cases illustrated in Figures \ref{fig:small_UC} and \ref{fig:large_UC}. The observed tuning values are largely consistent with the results presented in Table \ref{tab:percentage_an} for $a/r_0 = 200$. The decrease in the maximum range of tuning compared to the semianalytic result is due to the range of movement along $y$ also being significantly shorter. As can be seen, in the latter case of $b \approx 2a$ translation along $y$ yields tuning close to the maximum possible when the wires in the two sublattices are aligned along $y$.}
 \label{tab:percentage_exp}
\end{table}

\section{Results and discussion}



In summary, our work comprehensively studies the tuning strategies for the plasma frequency of wire-medium-based metamaterials. Assuming thin wires, we develop a semi-analytical theory predicting the plasma frequency for various wire lattices which is in good agreement both with the results of full-wave numerical simulations and experiments.

Quite importantly, we demonstrate that the plasma frequency can be flexibly reconfigured by up to 16\% while keeping the volume of the metamaterial practically unchanged. Such tuning appears to be remarkable for the entire field of tunable and reconfigurable metamaterials~\cite{Lapine2009,Gorkunov2008,Boardman2010}.

Hence, our work bridges the engineering world of metamaterials with the fundamental axion search experiments paving a way towards tunable wire-medium-based haloscopes for axion searches. High tunability reported here is one of the key ingredients needed for the success of the search experiments which may initiate further improvements in dark matter detection.


\section{Acknowledgements}

Analytical studies were supported by Priority 2030 Federal Academic Leadership Program. The experimental work was performed under support by the National Science Foundation, Grant No. NSF PHY220956.


\bibliographystyle{bibi}
\bibliography{bibliography}
\end{document}